\documentclass[titlepage,12pt]{article}

\begin{document}

\begin{titlepage}

\vskip .6in

\begin{center}
{\Large {\bf   Kappa-Symmetry of Green-Schwarz Actions in Coset 
Superspaces  }}

\end{center}

\normalsize
\vskip .6in

\begin{center}

{\bf I. N. McArthur}\footnote{E-mail: mcarthur@physics.uwa.edu.au}
\par \vskip .1in \noindent
 {\it Department of Physics, The University of Western Australia}\\
{\it Nedlands, W.A. 6907. Australia } 

\end{center}
\vskip 3cm

\begin{center}
{\large {\bf Abstract}}\\
\end{center}

The nature of $\kappa$-symmmetry transformations is examined for 
$p$-branes embedded in
 a class of coset superspaces $G/H,$ where $G$ is an appropriate supergroup 
and $H$ is the Lorentz subgroup. It is shown that one of the conditions 
$\delta Z^M \, E_M{}^a = 0$ which characterizes $\kappa$-symmetry 
transformations arises very naturally if they are implemented in 
terms of a right action of a subgroup of the supergroup $G$ on the 
supergroup elements which represent the coset. Unlike the global left 
action of $G$ on $G/H$ (which gives rise to supersymmetry on the coset 
superspace), there is no canonically defined right action 
of $G$ on $G/H.$ However, an interpretation of this right 
action involving an enlargement of the isotropy group from the 
Lorentz subgroup to a subgroup of $G$ with generators which include 
some of the fermionic generators of $G$ is suggested. Closure of the 
generators of this larger subgroup under commutation leads to the 
usual ``brane scan'' for $p$-branes which have bosonic degrees of 
freedom which are worldvolume scalars.

\vspace{9cm}
\noindent

\end{titlepage}

\section{Introduction}

Kappa symmetry is a local fermionic symmetry of covariant Green-Schwarz 
actions \cite{GS} for $p$-branes. It is a generalization of local 
fermionic symmetries discovered for massive \cite{Azc} and massless \cite{Siegel}
 superparticle actions.
The presence of the $\kappa$-symmetry in the Green-Schwarz action for 
a $p$-brane
ensures that only half of the supersymmetry generators 
are realized nonlinearly, as required for low energy effective 
actions describing the long wavelength excitations of BPS solitons 
\cite{Polch1,Polch2}. 
  The standard 
procedure for forming $\kappa$-symmetric Green-Schwarz actions 
in flat superspace is the ``nonlinear sigma model approach'' 
\cite{Mezincescu,AT}. The kinetic term  is a natural 
generalization of the Nambu action, to which is added a Wess-Zumino 
term formed 
by integrating a closed form in superspace over an embedded manifold 
whose boundary is the worldvolume of the $p$-brane\footnote{Only 
$p$-branes which have  bosonic degrees of freedom which are 
worldvolume scalars
will be considered in this paper.}. Global 
supersymmetry  of the action is ensured by constructing both pieces 
of 
the action in terms of the vielbein on flat superspace. 
 The $\kappa$-symmetry is then a 
field dependent local fermionic transformation for which the 
variations of the  the kinetic term and the Wess-Zumino term  can be 
made to 
cancel by appropriate choice of the normalization of the Wess-Zumino 
term. This procedure has recently been extended to the case of 
superspace backgrounds which are more general cosets than flat 
superspace \cite{Tseytlin2}.

Progress has been made in achieving a geometric understanding of 
kappa 
symmetry using the formalism  of super-embeddings 
\cite{STV}, based on doubly supersymmetric approaches to superstrings 
which incorporate both manifest worldvolume and spacetime 
supersymmetries \cite{Gates} 
(for a recent review of superembeddings including a 
comprehensive list of references, see \cite{Sorokin}).
The worldvolume of 
the $p$-brane is extended to a supermanifold which is embedded into 
the 
target superspace in such a way that the odd subspace of the 
worldvolume tangent space is a subset of the odd subspace of tangent 
space to the target superspace.
In such a formulation, the fermionic kappa symmetry can be 
related to the worldvolume supersymmetry.

 It has been noted in the past that 
in flat superspace, the $\kappa$-symmetry transformation of the 
bosonic superspace coordinate is like a ``wrong-signed'' 
supersymmetry transformation \cite{GSW}. In terms of the construction of 
flat superspace as the left coset $G/H$ of the super-Poincar\'e group $G$ with 
the Lorentz subgroup as the isotropy group $H$  \cite{Salam},
and in which supersymmetry transformations of the superspace coordinates 
are achieved by  the canonically defined left 
action of the superalgebra on the coset superspace,  
this is an indication that $\kappa$-symmetry transformations 
should be related to the {\em right} 
action of the superalgebra on the same coset. In 
the Hamiltonian formulation for $p$-branes embedded in flat superspace,
 de Azc\'arraga, 
Izquierdo and Townsend \cite{Towns6} showed that the fermionic 
constraints (some of which are first class, corresponding to the 
existence of local $\kappa$-symmetry) are analogous to  supercovariant 
derivatives in superspace, which generate right supertranslations.  Since 
left and right supertranslations commute, this provides a means to 
implement fermionic constraints which are consistent with the global 
supersymmetry generated by left supertranslations. 
 More recently Bars, 
Deliduman and Minic \cite{Bars}  have  implemented  
$\kappa$-symmetry transformations for a number of  {\em non-flat} 
coset superspaces in terms of a local right action.

The aim of  this paper is to point out  that for a general class of coset 
superspaces, $\kappa$-symmetry transformations do indeed have a natural 
interpretation in terms of a local right action, as opposed to the global 
left action which gives rise to the supersymmetry. In particular, 
after a review of the relevant properties of coset spaces in \S 2 
and a discussion of relevant coset superspaces in the first part of \S 3, it will 
be shown that the criterion
$$\delta Z^M \, E_M{}^a = 0$$ 
which characterizes $\kappa$-symmetry transformations in curved superspace 
\cite{Witten1,Gris1}
is very naturally implemented for coset superspaces $G/H$ 
in terms of a local right action of a part of the supergroup $G$. Although 
only the left action of the supergroup $G$ on the coset superspace $G/H$ is 
canonically defined (this is the usual nonlinear realization of $G$ on 
$G/H,$ which is independent of a particular parameterization of the 
coset), it is argued that the local right action associated with the 
$\kappa$-symmetry transformations is also well-defined, and should be 
interpreted as giving rise to an enlargement of the isotropy group $H$ from 
the Lorentz subgroup to a larger subgroup $\tilde{H}$ which includes 
fermionic generators. This is 
consistent with the role of $\kappa$-symmetry in 
covariant Green-Schwarz actions, which is to ensure that although there appear 
to be Goldstone fermions associated with all the supersymmetry generators, 
in fact only half of them are physical, meaning that only half of the 
supersymmetry is realized nonlinearly, with the rest of the supersymmetry 
being realized linearly on the physical degrees of freedom. In \S 4, 
this interpretation is expanded to include the worldvolume 
diffeomorphisms, which also eliminate some of the apparent Goldstone 
degrees of freedom in covariant Green-Schwarz actions. It is argued in 
\S 5 that consistency of this picture requires that the generators of the 
right actions 
associated with the $\kappa$-symmetry transformations and the worldvolume 
diffeomorphisms must form a closed superalgebra, from which emerge the 
usual restrictions on relationship between the dimension  of the 
worldvolume of a 
$p$-brane and the dimension of the spacetime into which it is embedded
 (the ``brane-scan''  \cite{Towns1}).

\section{Some Preliminaries}

We begin by reviewing relevant details of coset spaces\footnote{ The 
cosets considered in this paper are always left cosets.} and 
nonlinear realizations. Consider a Lie algebra $G$ with generators 
$T_{\cal{A}}$ 
satisfying the (possibly graded) commutation relations
$$ [T_{\cal{A}}, T_{\cal{B}}] = i f_{{\cal A} {\cal B}}{}{}^{{\cal 
C}} \, T_{\cal{C}}.$$
Let $T_{\bar{I}}$ denote the  generators of a subalgebra $H$ of 
$G$ (the ``unbroken'' generators), and 
$T_{A}$ denote the remaining  
generators of $G$ (the ``broken'' generators), and assume that the 
$T_A$ provide a representation of $H$ under commutation\footnote{ 
The index conventions to be used are as follows:  indices $\bar{I}, \bar{J}, 
\bar{K}, \ldots$ are used for the generators of $H;$ indices 
$A,B,C,\ldots$ are used for generators of $G$ not in $H;$ and 
$M,N,\ldots$ are used for coordinates on $G/H.$}.
 A ``slice'' through $G$ locally 
isomorphic to $G/H$ can be parameterized in the form 
\begin{equation} g(Z) = \exp (i Z^M \delta_M{}^{A} T_A),
\label{slice}
\end{equation}
with the $Z^M$ furnishing local coordinates on $G/H.$ There is a canonical global 
left action of $G$ on $G/H$ under which 
the point with coordinates $Z$ is mapped by fixed $g \in G$ into the 
point with coordinates $Z'$ determined by
\begin{equation}
 g . g(Z) = g(Z') . h(Z,g).
\label{leftaction}
\end{equation}
Here, $ h(Z,g)$ is the compensating $H$-transformation required to
bring $ g . g(Z)$ back onto the slice (\ref{slice}) isomorphic to $G/H.$ The 
realization of the group $G$ on the 
coordinates $Z$ is nonlinear.

The Cartan one-form $g(Z)^{-1} d g(Z)$ defined on $G/H$ takes values 
in 
the complexification of the Lie algebra of $G$ and can be decomposed 
as
\begin{equation}
g(Z)^{-1} d g(Z) = i\, dZ^M \, E_M{}^{A}(Z)\,T_{ A} +
 i\, dZ^M \, \Omega_M{}^{\bar{I}}(Z)\,T_{ \bar{I}} .
 \label{Cartan1}
 \end{equation}
These forms are invariant the global  left action of $G$ on 
$G/H$ 
since $d(g . g(Z)) = g . dg(Z)$ if $g$ is constant. Under the 
local right action $ g(Z) \rightarrow g(Z) . h(Z)$ with 
$h(Z) = e^{i \epsilon^{\bar{I}}(Z) T_{\bar{I}}}  
\in H,$ the forms 
$E^A(Z) = dZ^M \, E_M{}^{A}(Z)$ transform as a vielbein 
on the tangent bundle to  $G/H,$  while $\Omega^{\bar{I}}(Z) = dZ^M \, 
\Omega_M{}^{\bar{I}}(Z)$ transforms as a connection on this bundle (see, 
for example, \cite{Strathdee}).
The Maurer Cartan equations 
$0 = d( g(Z)^{-1}dg(Z)) + g(Z)^{-1}dg(Z) \wedge g(Z)^{-1}dg(Z)$
can be written in the form
\begin{eqnarray}
- \frac12 E^C(Z) \wedge E^B(Z)  f_{B 
C}{}{}^A &=& d E^A(Z) + \Omega^{\bar{I}}(Z) \wedge 
E^B(Z)  f_{B \bar{I}}{}{}^A  \nonumber \\
- \frac12 E^C(Z) \wedge E^B(Z)  f_{B
C}^{}{}^{\bar{I}} &=& d \Omega^{\bar{I}}(Z) + \frac12 \Omega^{\bar{K}}(Z)
 \wedge \Omega^{\bar{J}}(Z)  
f_{\bar{J}\bar{K}}{}{}^{\bar{I}},
\label{Maurer-Cartan}
\end{eqnarray}
so that the left-hand sides of these expressions are respectively the torsion and 
curvature of the coset space.

Replacing $d$ by $\delta$ in (\ref{Cartan1}),  an infinitesimal 
variation 
$\delta 
Z^M$ of the coordinate $Z^M$ on $G/H$ will induce a  variation 
$\delta g(Z)$ of the representative group element $g(Z)$  which 
satisfies 
\begin{equation}
g(Z)^{-1} \delta g(Z) = i \, \delta 
Z^M \, E_M{}^{A}(Z) \, T_{A} + i \, \delta 
Z^M \, \Omega_M{}^{\bar{I}}(Z) \, T_{\bar{I}}.
\label{delta}
\end{equation}

Consider now  the {\em right} 
action of the group $G$ on the representatives\footnote{Note that 
there is no canonical {\em right} action of the group $G$ on the 
(left) coset 
space $G/H,$ unlike the case of the {\em left} action of the group. 
The right action considered here is in general dependent on the choice of the 
slice $g(Z).$}  $g(Z)$ of the points in 
the coset $G/H.$ In particular, consider
 $$g(Z) \rightarrow g(Z) . e^{i( v^A (Z) T_{A} + w^{\bar{I}} (Z) T_{\bar{I}})}.$$  
Since the 
 new group element will lie on the $H$-orbit of some point on the 
 slice, one has
\begin{equation}
 g(Z) . e^{i( v^A (Z) T_{A} + w^{\bar{I}} (Z) T_{\bar{I}})} = g(Z') . h(v, w, Z)
 \label{localright}
 \end{equation}
 for some $ h(v, w, Z) \in H.$
In infinitesimal form one obtains
$$ g(Z') = g(Z). \left( 1 + i v^A (Z) T_{A} + i w^{\bar{I}} (Z) T_{\bar{I}}) \right) 
. \left( 
1 - i \phi^{\bar{I}}(v, w, Z) T_{\bar{I}} \right),$$
where $\phi^{\bar{I}}(v, w, Z)$ is linear in $v^A(Z)$ and $w^{\bar{I}}(Z).$
This yields 
\begin{equation}
 g(Z)^{-1} \delta g(Z) = i v^A(Z) T_{A}  + i \left( w^{\bar{I}}(Z) - 
 \phi^{\bar{I}}(v, w, Z) \right) \, T_{\bar{I}}.
\label{right}
\end{equation}
Comparing  (\ref{delta}) and (\ref{right}),  
\begin{equation}
 v^{ A}(Z) = \delta Z^M\, E_{M}{}^{A}(Z)
\label{infright}
\end{equation}
which allows $\delta Z^M$ to be computed given  $v^{A}(Z).$ The 
remaining relation
\begin{equation}
 \delta Z^M \, \Omega_M{}^{\bar{I}}(Z) =  w^{\bar{I}}(Z) - \phi^{\bar{I}}(v, w, Z)
\label{inf1right}
\end{equation}
determines the compensating Lorentz transformation $\phi^{\bar{I}}.$

\section{Coset Superspaces}

The analysis of $\kappa$-symmetric Green-Schwarz actions for 
$p$-branes in an generic ten or eleven dimensional superspace 
backgrounds \cite{Gris1,Plefka} is complicated by the fact that there
is no general prescription for the construction of the superfields 
which
correspond to a  given bosonic background \cite{Tseytlin1}. 
However, the fact that flat superspaces can be considered as cosets 
of 
the super-Poincar\'e group with the Lorentz subgroup as the stability 
group \cite{Strathdee2} allows a geometric formulation 
of the 
Green-Schwarz action in terms of  a nonlinear sigma 
model on the coset space \cite{Mezincescu,Polch2,Towns2}. Recently, 
this construction has been extended to the case of superspaces whose 
bosonic parts correspond to the near horizon geometries of various 
$p$-brane solitons, which are of the form $AdS_{p+2}\times 
S^{d-p-2}.$ The 
 corresponding superspaces can be realized as coset 
spaces $G/H$ for some choice of supergroup $G$ and bosonic stability 
group $H$, 
particular cases of interest being
\begin{eqnarray}
AdS_5 \times S^5 &\cong & SU(2,2|4)/\left(SO(1,4)\times SO(5)\right) 
\nonumber \\
AdS_4 \times S^7 &\cong & OSp(8|4)/\left(SO(1,3)\times SO(7)\right) 
\nonumber \\
AdS_7 \times S^4 &\cong & OSp(6,2|4)/\left(SO(1,6)\times 
SO(4)\right).
\label{coset}
\end{eqnarray}
These correspond respectively to the $D3$ brane in $d=10$ IIB 
supergravity \cite{Tseytlin2}, the $M2$ 
brane in $d=11$ supergravity, and the $M5$ 
brane in $d=11$ supergravity \cite{Plefka2,Claus}.

The  coset superspaces $G/H$ above are based on supergroups
$G$  with generators which will be denoted $P_{a},$ 
$Q_{\alpha}$ and $J_{ab}.$  The $J_{ab}$ are generators of a Lorentz 
subgroup $H,$ under which the even generators $P_{a}$ 
transform as a vector and the odd generators $Q_{\alpha}$ transform as 
spinors:
\begin{eqnarray}
\, [J_{ab}, P_c] &=& \frac{i}{2} \left( \eta_{ac}\, \delta_b{}^d - \eta_{bc} 
\, \delta_a{}^d \right) \, P_d \nonumber \\
\, [J_{ab}, Q^{\alpha}] &= & \frac{i}{8}\, [\Gamma_a, 
\Gamma_b]^{\alpha}{}_{\beta}\, Q^{\beta}.
\label{Lorentz}
\end{eqnarray}
The remaining (anti)commutation relations for the Lie 
algebra of the supergroup $G$ will not be  specified, 
except to require that the algebra has a contraction to the usual 
(possibly extended) supersymmetry algebra, as a result of which the 
anticommutator of  odd generators takes the form
\begin{equation}
\{ Q_{\alpha}, Q_{\beta} \} = -2 (C \Gamma^a)_{\alpha \beta}\, P_a + 
if_{\alpha \beta}{}^{ab}J_{ab}.
\label{flatSS}
\end{equation}
 This structure encompasses the 
super-Poincar\'e  group appropriate to the construction of flat 
superspaces, as well as  the  more general supergroups in the coset 
superspaces  (\ref{coset}). The coset superspaces   can be 
parameterized in the form 
$$g(Z) = e^{i(x^{m}\delta_m{}^a P_{a} + \bar{\theta}Q )} \equiv e^{i 
Z^M \delta_M{}^A T_A},$$
where $Z^M  = (x^m, \theta^{\alpha}),$ and $T_A = (P_a, Q_{\alpha})$ are 
the ``broken'' generators (and $T_{\bar{I}} = (J_{ab})$ are the 
``unbroken'' generators).

Evidence is now provided for 
the identification of the $\kappa$-symmetries of the Green-Schwarz 
actions in these  coset superspaces $G/H$ with a {\em right} 
action\footnote{The term ``action'' appears twice in this sentence, 
once in the context of the integral of a Lagrangian, and once in the 
context of transformations on a group manifold generated by the group 
structure; in future use of this term, it should be clear from the 
context which of 
these is applicable.} of the 
the supergroup $G.$ 
The canonically defined 
global action of an element $g \in G$ on the coset superspace is 
determined the left action $g.g(Z),$ as in (\ref{leftaction}). A
Green-Schwarz action is automatically invariant under this group 
action since
it is constructed from the Cartan forms $g(Z)^{-1}dg(Z).$ Consider 
instead the local transformation 
$$g(Z) \rightarrow g(Z') = g(Z)\, e^{i v^A(Z)T_A}\,e^{-i 
\phi^{ab}(v,Z)J_{ab}} ,$$
with $v^AT_A = v^a P_a + \bar{v}_{\alpha} Q^{\alpha}.$
As noted earlier, the right action of the group on representative of 
cosets does not descend to a canonically defined transformation on 
$G/H:$  the transformation $Z \rightarrow Z'(Z)$ determined 
by (\ref{localright}) is dependent on the choice of slice $g(Z).$ In 
infinitesimal form, the relationship between $v^{A}(Z)$ and 
$\delta Z = Z' - Z$ is given by (\ref{infright}),
  $$  v^A(Z) = \delta Z^M \, E_M{}^{A}(Z),$$
where $E_M{}^{A}(Z)$ is the vielbein on the coset superspace. Thus, setting 
$v^a(Z) = 0$ corresponds to the condition 
$$0 = \delta Z^M \, E_M{}^{a}(Z).$$
{\em This is one of the criteria which characterize a 
$\kappa$-transformation on the worldvolume of a $p$-brane embedded in 
a curved superspace} \cite{Witten1,Gris1}. It 
determines the transformation $\delta x^m$ of the even superspace
coordinates of the embedded worldvolume in terms of $\delta  
\theta^{\alpha},$ the 
variation of the odd coordinates.

So the condition that 
$\delta Z^M \, E_M{}^{a}(Z)$ vanishes is naturally implemented in 
terms of 
a local right action depending only on the fermionic 
generators of the supergroup\footnote{Of course, 
there are additional criteria which a $\kappa$-symmetry 
transformation 
must fulfill, as will be discussed shortly. In 
particular, only a subset of the supersymmetry generators  is involved in a 
$\kappa$-symmetry transformation, and this subset depends 
 on the point on the embedded worldvolume. }.
As there is no canonically defined right action 
of $G$ on $G/H,$  the implementation of a 
$\kappa$-transformation in the above fashion would seem to depend on 
the  parameterization of the coset superspace. However, this is not 
true provided the right action is  a {\em local
 symmetry}\, ; it is then a reflection of the presence of 
unphysical degrees of freedom which can be eliminated by a choice of 
gauge for the local symmetry.

The latter principle is well known in the standard construction of a 
sigma model (not necessarily supersymmetric) with a target space 
which is a coset space. Generically  \cite{Nicolai}, 
one starts with an action of the 
form
$$ S =  \int dx \, (g^{-1}\partial_{m}g)^{A} \, 
(g^{-1}\partial^{m}g)^{B}\, \delta_{AB}$$
where $g(x) = \exp^{i ( \xi^{A}(x) T_{A} + \xi^{\bar{I}}(x) T_{\bar{I}})} \in G,$ 
with  $T_{\bar{I}}$ 
the generators of $H,$ and where $\delta_{AB}$ is a $H$-invariant 
metric. The sigma model possess a 
local symmetry under the right action of $H$ since 
$(g^{-1}\partial_{m}g)^{A}$
transforms linearly under such transformations (assuming the generators 
$T_A$ provide a representation of $H$ under commutation with the $T_{\bar{I}}).$
This gauge symmetry 
can be fixed 
by setting $\xi^{\bar{I}}$ to zero, giving a parameterization of $G/H$ in a 
physical gauge in  
terms of the slice $g(\xi) = e^{i  \xi^{A} T_{A}}.$ Since it is 
constructed from the Cartan form $g^{-1}dg,$ the action $S$ also has a 
global symmetry under the left action of $G.$ In particular,  this 
global left action gives a linear realization of the subgroup $H$
 on the physical degrees of freedom $\xi^{A},$ while 
the rest of the generators of $G$ are realized nonlinearly.  Even 
in the  physical gauge, there is still a symmetry under the 
local right action of $H.$

In the case of 
$p$-brane actions in coset superspaces $G/H,$ a local symmetry under 
the right action of a Lorentz subgroup $H$ has been fixed
in choosing a ``slice''  $g(Z)$ which is parameterized in the form 
$g(Z) = e^{i(x^m \delta_m{}^a P_a + \bar{\theta}Q)}.$ 
If the Wess-Zumino term is appropriately normalized, then 
the action possesses a $\kappa$-symmetry, which,  as argued above, 
takes the form 
of a local right action generated by a subset of the supersymmetry 
generators. Although this right action is not well-defined on $G/H,$ 
it suggests that the theory really describes degrees of freedom on a 
smaller coset space $G/\tilde{H},$ where $\tilde{H} \supset H$ 
includes an appropriate subset of the fermionic generators. In this 
case the  right action {\em is} well-defined, independent of the 
parameterization of $G/H$ with which the construction starts. 
Further, under the left action of $G,$ the fermionic generators in 
$\tilde{H}$ will be realized linearly on the physical degrees of 
freedom, while the remaining fermionic generators will still be nonlinearly 
realized.  This is precisely the requirement for the effective action 
describing the low energy excitations of a BPS state which preserves 
only
half the supersymmetry. It is this interpretation of $\kappa$-symmetry 
which we wish to pursue.
 It should be pointed out that the 
situation is actually more subtle, because the subgroup  $\tilde{H}$ 
varies from point to point of the embedded worldvolume.

To illustrate the above, we review   $p$-branes in 
flat superspace, which is a coset superspace $G/H$ with $G$ a 
super-Poincar\'e group and $H$ its Lorentz subgroup. The coset is 
parameterized in the form $g(Z) = e^{i(x^aP_a + \bar{\theta}Q)},$  
and the Cartan form \begin{eqnarray}
g(Z)^{-1}dg(Z) & = & i \Pi^a P_a + i d\bar{\theta}Q \nonumber \\
&=& i \left( dx^a - i (\bar{\theta}\Gamma^a d \theta ) \right) P_a +
i\, d\bar{\theta}Q
\label{flatcartan}
\end{eqnarray}
  provides a 
vielbein on flat superspace which is invariant under global 
supersymmetry transformations
$$  \delta x^a =  i (\bar{\epsilon} \Gamma^a \theta ),\,\,\,\,\,\,
\delta \theta = \epsilon .$$
In infinitesimal form, the local right action 
$g(Z) \rightarrow g(Z) \, e^{i \bar{\epsilon} Q}$ induces the 
transformations 
\begin{equation}
\delta x^a = - i (\bar{\epsilon} \Gamma^a \theta ),\,\,\,\,\,\,
\delta \theta = \epsilon
\label{flatkappa}
\end{equation} (a compensating Lorentz transformation is not 
required in flat superspace). The vielbein transforms as
$$ \delta \Pi^a = -2i\, (\bar{\epsilon} \Gamma^a d\theta), \,\,\,\, \delta 
\, d\theta = d \epsilon.$$ The  Lorentz invariant metric $\eta_{ab}$ 
on the bosonic part of flat superspace can be pulled back to the $(p+1)$-dimensional 
worldvolume of the $p$-brane to yield a worldvolume metric
$$G_{ij} = \frac{\partial Z^M}{\partial \sigma^i}
\frac{\partial Z^N}{\partial \sigma^j} \Pi_M{}^a \Pi_N{}^b 
\eta_{ab} \equiv \Pi_i{}^a \Pi_j{}^b \eta_{ab},$$
where $\sigma^i$ are coordinates on the  worldvolume. The kinetic 
term $S_0$  in the $p$-brane action is a generalization of the massive 
spinning particle action, being proportional to the volume of the 
woldvolume with respect to the induced metric:
$$S_0 = \int d^{(p+1)} \sigma \, \sqrt{\det G}.$$
Since $\Pi^a$ can be written in the form
$$ i \Pi^a P_a = g(x,\theta)^{-1} D g(x,\theta),$$
where $Dg = d g - i g \,(d\bar{\theta}Q)$ is a covariant derivative of 
$g,$  $ \Pi^a$ transforms linearly under the local right action 
$g(Z) \rightarrow g(Z) \, e^{i \bar{\epsilon} Q}.$ However, $G_{ij}$ 
is not invariant under this transformation because $\eta_{ab}$ is 
an invariant metric only for the Lorentz subgroup of 
the super-Poincar\'e group, with the result 
that the variation of the kinetic term in the $p$-brane action is  
\begin{equation}
 \delta \int d^{(p+1)} \sigma \, \sqrt{\det G} = -2 i
\int d^{(p+1)} \sigma \, \sqrt{\det G} \, (\bar{\epsilon} 
\Gamma^i \partial_i \theta).
\label{deltaS0}
\end{equation}
Here, the $\Gamma_i$ are worldvolume gamma matrices 
 obtained by pulling back the spacetime gamma matrices,
$$\Gamma_i =  \frac{\partial Z^M}{\partial \sigma^i} \Pi_M{}^{a} 
\Gamma_a.$$
In the usual manner for the construction of sigma models with degrees 
of freedom on a coset space \cite{Nicolai}, if 
 $\Pi_i{}^a$ and $\Pi_j{}^b$ {\em could} be contracted with an 
invariant metric for the full super-Poincar\'e group, then an action could 
be constructed with a local fermionic symmetry.  It would describe
degrees of freedom living in the space of left cosets of flat superspace 
under the local fermionic transformation 
$g(Z) \rightarrow g(Z) \, e^{i \bar{\epsilon} Q}.$ 
However, an appropriate invariant metric does not exist. 

As is now well known, it {\em is} possible to obtain partial invariance with 
respect to the local right action. 
The matrix\footnote{We use the 
convention $d\sigma^{i_0} \wedge d\sigma^{i_1} \wedge \cdots 
d\sigma^{i_p} =$ $d^{(p+1)}\sigma \, \sqrt{\det G}\,
 \epsilon^{i_0 i_1 \cdots i_p}.$ Also, a rotation to a Euclidean 
 metric has been made for convenience.}
$$ \Gamma = \frac{i^{[\frac{p+1}{2}]}}{(p+1)! }\,
 \epsilon^{i_0 i_1 \cdots i_p} 
\Gamma_{i_0} \Gamma_{i_1} \cdots \Gamma_{i_p}$$ has the property 
$\Gamma^2 = 1$, and commutes with 
the worldvolume gamma matrices $\Gamma_i$ 
when $p$ is even and anticommutes with them when $p$ is odd. The 
 generators $J_{ij} = - \frac{i}{8} [\Gamma_i, 
\Gamma_j]$ therefore yield a reducible representation of the worldvolume 
Lorentz group, and the projection operators
 $P_{\pm} = \frac12 (1 \pm \Gamma)$ provide a 
decomposition of this reducible representation into two parts. If it 
is possible to find a contribution $S_{WZ}$ to the action such that 
under the local transformations
$$\delta (S_0 + S_{WZ}) = - 4 \, i 
 \int d^{(p+1)} \sigma \, \sqrt{\det G} \, (\bar{\epsilon} 
P_- \,\Gamma^i \partial_i \theta),$$ then $S_0 + S_{WZ}$ will be 
invariant under transformations 
$g(Z) \rightarrow g(Z) \, e^{i \bar{\epsilon} Q}$ 
with $\bar{\epsilon} = \bar{\kappa} 
P_+.$ These are the $\kappa$-symmetry transformations, which allow 
half of the fermionic degrees of freedom to be gauged away.

A contribution $S_{WZ}$ to the $p$-brane action with the required 
transformation property under the local right action and which is 
invariant  under global supersymmetry (or a global left action) exists 
only for flat superspaces in which the ($p$+2)-form 
$$ h = i^{[\frac{p+1}{2}]}\, \frac{i}{p! }\, \Pi^{a_1} \wedge \Pi^{a_2} \wedge \cdots \wedge \Pi^{a_p} \, (d 
\bar{\theta}\, \Gamma_{a_1} \Gamma_{a_2} \cdots \Gamma_{a_p} d\theta)$$ 
is closed; this leads to the ``brane-scan'' restricting the 
relationship between the dimension of the worldvolume and the 
dimension of the spacetime in which it is embedded \cite{Towns1}.
The closure of this form is equivalent to the integrability of 
\begin{equation}
 \delta S_{WZ} = 2 \, i \int d^{(p+1)} \sigma \, \sqrt{\det G} \, 
 (\bar{\kappa}\, 
\Gamma\,\Gamma^i \partial_i \theta),
\label{varWZ}
\end{equation} 
in the sense that only in these 
cases is there a local quantity $S_{WZ}$ whose variation yields the 
right hand side of (\ref{varWZ}). This is similar to the Wess-Zumino 
consistency condition for anomalies \cite{WZconsistency}.

In this case, 
the $p$-brane action is
$$ S_0 + S_{WZ} = \int d^{(p+1)} \sigma \, \sqrt{\det G} + \int \phi^{*} b,$$
 $\phi$ denotes the 
embedding of the worldvolume into superspace, and $b$ is a ($p$+1)-form 
defined by $db = h.$ To see how this expression for the Wess-Zumino term  arises, note that 
$ \delta S_{WZ}$ can be written in the form 
\begin{eqnarray*}
 \delta S_{WZ} &=& i^{[\frac{p+1}{2}]}\, \frac{2\, i}{p! }\int d \sigma^{i_1} \wedge \cdots \wedge d 
\sigma^{i_{p+1}}  \, \left( \bar{\kappa} \Gamma_{i_1} \cdots \Gamma_{i_{p}} 
\partial_{i_{p+1}} \theta \right) \\
&=& i^{[\frac{p+1}{2}]}\, \frac{2 \, i}{p! }  \int \phi^* \, \Pi^{a_1} \wedge \cdots \wedge \Pi^{a_p} 
\wedge \left( \bar{\kappa} \Gamma_{a_1} \cdots \Gamma_{a_p} d\theta \right).
\end{eqnarray*}
If we consider a path $g(t) = e^{i(x^aP_a + t\bar{\theta}Q)}$ in 
the {\em fermionic} directions on the coset superspace, then this is 
equivalent to 
\begin{equation}
\frac{d}{dt} S_{WZ}(t) = i^{[\frac{p+1}{2}]} \, \frac{2 \, i}{p! } \int \phi^* \, \Pi^{a_1}(t) \wedge \cdots 
\wedge \Pi^{a_p}(t) 
\wedge \left( \bar{\nu}(t) \Gamma_{a_1} \cdots \Gamma_{a_p} \, t \, d\theta 
\right)
\label{*}
\end{equation}
where 
$$ g(t)^{-1}dg(t) = i \Pi^a(t) P_a + i t \,d\bar{\theta}Q =  
i \left( dx^a - i t^2 (\bar{\theta}\Gamma_a d\theta)\right) P_a +
 i t \,d\bar{\theta}Q $$
and\footnote{Note that 
there is no $P_a$ term in $g(t)^{-1}\frac{d}{dt}g(t)$ for the particular choice of path $g(t)$ given 
because $dt \, (\bar{\theta}\Gamma^a \theta) = 0;$ for a more general path 
$\theta(t),$ there would be such terms, in which case it would also be 
necesary to include an integration of
 the variation of $S_{WZ}$ under the bosonic 
part of the infinitesimal
right action.}
$g(t)^{-1}\frac{d}{dt}g(t) = i(\bar{\nu}(t) Q).$  
The reason that a derivative with respect 
to $t$ can be considered as an infinitesimal right action with 
parameter $\bar{\nu}(t)_{\alpha}$ is that $\frac{d}{dt} g(t)$ can 
be written in the form $g(t) . 
(g(t)^{-1}\frac{d}{dt}g(t)), $ mimicing an infinitesimal
 local right action. Equation
 (\ref{*}) can be 
integrated\footnote{The approach of integrating the variation of the 
Wess-Zumino term  was used by Aganagic et 
al  to construct D-$p$-brane Wess-Zumino terms in 
\cite{Schwarzetal}.}  to yield
$$ S_{WZ} = i^{[\frac{p+1}{2}]} \, \frac{2 \, i}{p! }\int_0^1 dt \int \phi^* \, \Pi^{a_1}(t) \wedge \cdots 
\wedge \Pi^{a_p}(t) 
\wedge \left( \bar{\nu}(t) \Gamma_{a_1} \cdots \Gamma_{a_p}\, t \, d\theta 
\right).$$ The $t$ integral on right hand side of this expression is the standard 
cohomological construction of a form $b$ with the property $db = h$ 
in the case where $h =  i^{[\frac{p+1}{2}]} \, \frac{i}{p! }\, \Pi^{a_1} \wedge \Pi^{a_2} \wedge \cdots \wedge
 \Pi^{a_p} \, (d \bar{\theta} \, \Gamma_{a_1} \Gamma_{a_2}
  \cdots \Gamma_{a_p} d\theta)$ is closed.

Note that although flat superspace can be constructed as a coset space of 
the super-Poincar\'e group by the Lorentz group, it is in fact itself 
a group, the super-translation group (as emphasized in \cite{AT}).
 So the left and right actions of the super-translation group 
on flat superspace are well-defined. This is reflected in the fact 
that no compensating Lorentz transformation is required in the 
definition of the right action of the super-Poincar\'e group on 
$g(Z) = e^{i(x^aP_a + \bar{\theta}Q)}.$ The  entire discussion 
of $p$-branes in flat superspace above can therefore be carried out without 
reference to cosets.  Nevertheless, it is useful to 
consider  flat superspace as a 
coset space, because it makes obvious the origin of the local Lorentz invariance of the $p$-brane 
action, and also because it points to the generalizations required 
for coset superspaces $G/H$ which are not groups and for which there is no 
naturally  defined right action of $G.$

\section{Incorporating Worldvolume Diffeomorphisms}

Returning to the case of $p$-brane actions in general coset superspace 
backgrounds, they are by construction also invariant under 
worldvolume diffeomorphisms. This ensures that there are only 
physical Goldstone modes corresponding to the generators $P_a$ of 
superspace translations which are broken by the embedding of the 
$p$-brane into superspace. The unphysical bosonic Goldstone degrees of freedom 
in the covariant Green-Schwarz action can be eliminated by a choice 
of gauge for the worldvolume diffeomorphisms. Since symmetry with 
respect to the worldvolume 
diffeomorphisms performs the same role as the $\kappa$-symmetry 
transformations, namely the elimination of unphysical Goldstone 
degrees of freedom, then if the above interpretation of the 
$\kappa$-symmetry transformations as enlarging the isotropy group in 
the construction of the coset superspace is correct, the worldvolume 
diffeomorphisms should also have an interpretation in terms of the 
local right action of a subgroup of the supergroup $G$ on the coset 
superspace $G/H.$ This is indeed the case, as we now demonstrate.

If $Z^M(\sigma)$ denotes the embedding of the $(p+1)$-dimensional 
worldvolume (with coordinates $\sigma^i)$ of a $p$-brane into the 
coset superspace, then since the spacetime coordinates are worldvolume 
scalars, they transform under an infinitesimal worldvolume 
diffeomorphism generated by the worldvolume vector field $v^i(\sigma) 
\, \frac{\partial}{\partial \sigma^i}$ as
$$ \delta_v Z^M (\sigma) = v^i(\sigma) \,  \frac{\partial Z^M (\sigma)}{\partial 
\sigma^i}.$$ Using (\ref{infright}) and (\ref{inf1right}), such a 
transformation of the spacetime coordinates is 
induced by a right action of the form
$$g(Z( \sigma )) \rightarrow g(Z'( \sigma )) = g(Z(\sigma )) \, 
e^{iv^A (\sigma ) T_A + i w^{ab}(\sigma) J_{ab}} \, e^{-i 
\phi^{ab}(\sigma)J_{ab}}$$ with
$$v^A(\sigma) = v^i(\sigma) \frac{\partial Z^M (\sigma)}{\partial 
\sigma^i} E_M{}^A(Z(\sigma))$$ and $$w^{ab}(\sigma) =
 v^i(\sigma) \frac{\partial Z^M(\sigma)}{\partial 
\sigma^i} \Omega_M{}^{ab}(Z(\sigma)) - \phi^{ab}(v,\sigma).$$ It is 
possible to choose $w^{ab}(\sigma)$ so that the compensating 
transformation vanishes, namely $w^{ab}(\sigma) =
 v^i(\sigma) \frac{\partial Z^M(\sigma)}{\partial 
\sigma^i} \Omega_M{}^{ab}(Z(\sigma)).$ In this case the diffeomorphism 
is generated in the form
\begin{equation}
  g(Z'(\sigma)) = g(Z(\sigma)) \, 
e^{iv^i(\sigma) (E_i{}^A(\sigma)T_A + \Omega_i{}^{ab}(\sigma)J_{ab})},
\label{diffeo2}
\end{equation}
where $E_i{}^A = \frac{\partial Z^M}{\partial 
\sigma^i} \, E_M{}^A$ and $\Omega_i{}^{ab} = \frac{\partial Z^M}{\partial 
\sigma^i} \, \Omega_M{}^{ab}$ are the pullbacks of the 
vielbein and connection to the worldvolume. This is very similar to the 
implementation of diffeomorphisms in superspaces in terms of so-called 
supergauge transformations \cite{wess2}.

To check this, computing $g(Z')^{-1} \frac{\partial}{\partial \sigma^i} g(Z') $ 
with $v^i(\sigma)$ infinitesimal yields (upon use of the expressions 
(\ref{Maurer-Cartan})
for the torsion and curvature of a coset superspace) the correct 
transformations
\begin{eqnarray}
\delta_v E_i{}^A(\sigma) &=& \frac{\partial v^j (\sigma)}{\partial 
\sigma^i} E_j{}^A(\sigma) +
v^j(\sigma) (\partial_j E_i{}^A(\sigma))   \nonumber \\
\delta_v \Omega_i{}^{ab}( \sigma) 
&=&\frac{\partial v^j(\sigma)}{\partial \sigma^i} 
\Omega_j{}^{ab}(\sigma) +
v^j(\sigma) (\partial_j \Omega_i{}^{ab}(\sigma))
\label{X}
\end{eqnarray}
for the pullbacks of the vielbein and connection. 

\section{The Algebra of $\kappa$-Transformations and Diffeomorphisms}

Both the $\kappa$-symmetry transformations and 
the diffeomorphisms are local symmetries which remove unphysical 
Goldstone degrees of 
freedom from the covariant Green-Schwarz action for a $p$-brane 
propagating in a coset superspace $G/H,$ leaving  physical 
Goldstone modes only for those generators of the supergroup $G$ which 
are broken by the embedding of the $p$-brane in superspace. In the 
left action of $G$ on $G/H,$ the broken 
generators are realized nonlinearly on the physical degrees of 
freedom, while the unbroken generators are realized linearly. As 
already mentioned, this is naturally achieved if the $\kappa$-symmetry 
transformations and the diffeomorphisms arise from a 
local right action of special linear combinations of the generators 
of the supergroup $G$ on the coset superspace 
representatives $g(Z(\sigma));$ {\em however, for this picture to be 
consistent, these linear combinations of generators must close under 
commutation and so generate a subgroup of the supergroup.} This will 
then have the effect of enlarging the isotropy group of the coset 
superspace to a subgroup $\tilde{H}$ 
which contains these linear combinations of  generators in addition to the 
Lorentz generators. The subgroup $\tilde{H}$ varies from point to 
point on the embedded worldvolume.

It has already been demonstrated in a general context that both the 
$\kappa$-symmetry transformations and the diffeomorphisms can indeed 
be realized in terms of a local right action. It thus remains to show that 
the corresponding generators form a subalgebra of the full 
superalgebra $G.$ In other words, it is necessary to  show
 that the generators of
$\kappa$-symmetry transformations, worldvolume diffeomorphisms and 
Lorentz transformations form a closed algebra under commutation if 
the above interpretation of these transformations as local right 
actions is to be consistent.

Before calculating the appropriate commutators, 
the form of $\kappa$-symmetry transformations in a general coset 
superspace must be specified.
As in flat superspace, they involve 
the pullback to the worldvolume of various structures on the coset 
superspace. These include the worldvolume metric, which is the 
pullback of the Lorentz invariant metric $\eta_{ab}:$ 
$$G_{ij}(\sigma) = E_i{}^a(Z(\sigma)) \,   E_j{}^b(Z(\sigma)) 
\,\eta_{ab},$$
where $E_i{}^A = \frac{\partial Z^M}{\partial \sigma^i} E_M{}^A.$
Similarly,  worldvolume gamma matrices are obtained by pulling back 
those defined in spacetime relative to the metric  $\eta_{ab}:$ 
$$\Gamma_i(\sigma) =  E_i{}^a(Z(\sigma)) \Gamma_a$$
satisfies $\{\Gamma_i(\sigma),\Gamma_j(\sigma)\} = 2 G_{ij}(\sigma).$
The  matrix
$$ \Gamma = \frac{i^{[\frac{p+1}{2}]}}{(p+1)!}\,
 \epsilon^{i_0 i_1 \cdots i_p} 
\Gamma_{i_0} \Gamma_{i_1} \cdots \Gamma_{i_p}$$ has the same 
properties as in the case of flat superspace, and the projection operators
 $P_{\pm} = \frac12 (1 \pm \Gamma)$ provide a 
decomposition of the spinor representation of the worldvolume Lorentz 
group provided by the worldvolume gamma matrices into two parts.
A related projection operator is the symmetric tensor $K^{ab}(\sigma)  = 
E_i{}^{a}(\sigma) G^{ij}(\sigma)E_j{}^{b}(\sigma),$ 
 satisfying $K_a{}^b K_b{}^c = K_a{}^c.$ It has 
the property  
\begin{equation}
E_i{}^{b}(\sigma)K_b{}^a(\sigma) = 
E_i{}^{a}(\sigma).
\label{K}
\end{equation}
If $\phi = E^a \phi_a$ is a bosonic one-form on the coset superspace, 
then $K_a{}^b(\sigma) \phi_b(\sigma)$ is the piece of $\phi_a$ which 
pulls back to the worldvolume, while $(1 - K)_a{}^b(\sigma) \phi_b(\sigma)$ 
has a vanishing pull-back. For the spacetime gamma matrices 
$\Gamma_a, $ this means that $\Gamma_b K_b{}^a(\sigma) $ pulls back to a 
linear combination of the worldvolume gamma matrices, $\Gamma^b 
K_b{}^a(\sigma) 
= \Gamma^i(\sigma) E_i{}^a(\sigma),$  
 while $\Gamma^b (1-K)_b{}^a(\sigma) $
consists of the gamma matrices which are ``transverse'' to 
the worldvolume gamma matrices (and therefore anticommutes with all 
of them). This will be important in what follows.

The $\kappa$-symmetry transformations for a $p$-brane embedded in flat 
superspace are of the form
$$ g(Z(\sigma)) \rightarrow g(Z(\sigma)) \, e^{i \bar{\kappa}(\sigma) 
P_+(\sigma) Q}.$$
Note that the transformation refers to a particular point $Z(\sigma)$ on the 
embedded worldvolume, and that the spinor  $\kappa^{\alpha}$ is an 
arbitrary function of $\sigma.$ 
In a curved coset superspace,  the point 
$g(Z(\sigma)) \, e^{i \bar{\kappa}(\sigma) 
P_+(\sigma) Q}$ no longer lies on the slice $g(Z) = e^{i Z^M \delta_M{}^A 
T_A},$ and so  a compensating Lorentz transformation is 
necessary if $\kappa$-symmetry transformations are to take this form.
 Alternatively, by the same method used in the case of 
diffeomorphisms, it is possible to implement the $\kappa$-symmetry 
transformation in such a manner that a compensating Lorentz 
transformation is not necessary, in which case the right action is of 
the form
\begin{equation}
g(Z(\sigma)) \rightarrow g(Z'(\sigma)) = 
 g(Z(\sigma)) \, e^{i \bar{\kappa}(\sigma) 
P_+(\sigma) Q + i (\bar{\kappa}P_+ C^{-1})^{\alpha} E_{\alpha}{}^N 
\Omega_N{}^{ab}(\sigma)J_{ab}}.
\label{kappa}
\end{equation}
In deriving this, it is necessary to use 
$$\delta_{\kappa} Z^M \Omega_M{}^{ab} = \delta_{\kappa} Z^M E_M{}^A E_A{}^N 
\Omega_N{}^{ab},$$
where $E_A{}^N$ is the inverse vielbein, as well as the fact that 
$\delta_{\kappa} Z^M E_M{}^{\alpha} =  (\bar{\kappa} P_+ C^{-1})^{\alpha}.$

In the appendix, it is shown that for infinitesimal worldvolume 
diffeomorphisms and $\kappa$-symmetry transformations, defined in 
terms of the right actions (\ref{diffeo2}) and (\ref{kappa}), the 
following commutation relations hold:
\begin{eqnarray}
\, [\delta_{v_1}, \delta_{v_2}] &=& \delta_{v_3}\nonumber \\
\, [\delta_v,\delta_{\kappa}] &=& \delta_{\kappa'},
\label{difkap}
\end{eqnarray}
with $v_3^i = v_2^j(\partial_j v_1^i) -  v_1^j(\partial_j v_2^i)$ and 
$\kappa' = - v^i \partial_i \kappa.$ The first commutator yields 
the expected algebra for worldvolume diffeomorphisms, and the second 
commutator is consistent with long-established properties of the 
algebra of  $\kappa$-symmetry transformations 
\cite{GS,Towns2}. 

The commutation 
relations (\ref{difkap}) are valid independent of the dimension of 
the spacetime formed by the coset superspace into which the $p$-brane 
is embedded. However, this is not true for the commutator 
$[\delta_{\kappa_1},\delta_{\kappa_2}]$ of a pair 
of $\kappa$-symmetry transformations.  As shown in the 
appendix, such 
a commutator yields a linear combination of a diffeomorphism 
(parameterized by the worldvolume vector field
 $v^i = -2 i (\bar{\kappa}_2 P_+ \Gamma^i P_{\pm} \kappa_1)),$ a 
Lorentz transformation and  another $\kappa$-symmetry 
transformation\footnote{The calculation in the appendix does not give the 
explicit form of the Lorentz transformation and the $\kappa$-symmetry 
transformation, although these could be extracted with a little more 
effort. Also note that 
conditions (i) and (ii) are only shown  
to be sufficient rather than necessary conditions for the algebra to close.} 
if the following conditions are satisfied: 

\hspace{1cm} (i) The dimension of the spacetime into which the $p$-brane is 
embedded is such that the gamma matrix identities 
$$0 = (C \Gamma_{a})_{\alpha (\beta} \, (C \Gamma^{a})_{\gamma 
\delta)}$$
for $p=1$ and 
$$0 = (C \Gamma_{a_1})_{(\alpha \beta} \, (C \Gamma^{a_1 a_2 \cdots 
a_p})_{\gamma \delta)}$$
for $p>1$ are true (the round brackets on spinor indices denote 
symmetrization). This yields the usual ``brane scan'' for 
$p$-branes whose bosonic degrees of freedom are worldsheet scalars 
\cite{Towns1}.

\hspace{1cm} (ii) For $p>1,$ the additional condition $0 = (P_- 
\Gamma^i)^{\alpha}{}_{\beta} E_i{}^{\beta}$ must also be satisfied. 

\noindent
Condition (ii) is one  of the equations of motion for a $p$-brane, 
namely  that which follows from 
\begin{equation}
\delta S = \int d^{(p+1)} \sigma \, \sqrt{\det G} \, (\delta Z^M) 
E_M{}^{\alpha} (C P_- 
\Gamma^i)_{\alpha \beta} E_i{}^{\alpha}
\label{action}
\end{equation}
under a variation $(\delta Z^M) 
E_M{}^{\alpha}$  of the $p$-brane action $S.$
This is  
consistent with the existence of a $\kappa$-symmetry form 
$(\delta_{\kappa} Z^M) E_M{}^{\alpha} = (\bar{\kappa} P_+ 
C^{-1})^{\alpha}. $ As discussed in the case of flat superspace,
 (\ref{action}) can be taken as a {\em 
definition} of the $p$-brane action \cite{Schwarzetal}, 
in that integrating this gives 
the usual $p$-brane action (with the kinetic term coming from the 
piece in $P_-$ proportional to $1$ and the Wess-Zumino term coming 
from the piece in $P_-$ proportional to $\Gamma ).$ So except for 
$p=1$ (a string), 
the algebra of diffeomorphisms and $\kappa$-symmetry 
transformations closes only on-shell. It is interesting that 
requiring the 
algebra of the $\kappa$-symmetry transformations, worldvolume 
diffeomorphisms and Lorentz transformations to close on itself 
yields a condition 
which can be used to construct the $p$-brane action.

\section{Conclusion}

The suggestion \cite{GSW,Towns6,Bars} that $\kappa$-symmetry transformations are  
related to the right action of the supergroup $G$ on a coset 
superspace $G/H$ has been examined for a special class of coset 
superspaces which are currently of interest  in studies of $p$-branes 
propagating in curved spacetimes. It has been shown that the  
condition 
$$\delta Z^M \, E_M{}^a = 0$$
which is accepted as one of the hallmarks of a $\kappa$-symmetry 
transformation for $p$-branes embedded in curved superspaces 
\cite{Witten1,Gris1}  
arises very naturally if the $\kappa$-symmetry transformation is 
implemented in terms of a right action of the supergroup $G$ on $G/H.$ 
Further, a physical interpretation of this right action in terms of an 
enlargement of the isotropy subgroup from the Lorentz subgroup of $G$ 
has been suggested. This is consistent with the role of both 
worldvolume diffeomorphisms and $\kappa$-symmetry transformations in 
covariant Green-Schwarz $p$-brane actions, namely in the elimination of 
unphysical Goldstone degrees of freedom. Requiring that the algebra of 
worldvolume diffeomorphisms, $\kappa$-symmetry transformations and 
Lorentz transformations closes on itself at each point of the embedded 
worldvolume leads to the usual ``brane scan'' for $p$-branes with 
worldvolume degrees of freedom which are scalars \cite{Towns1}.
 In the case $p>1,$ 
closure of the algebra also requires that one of the usual equations of 
motion for a $p$-brane be true. It was pointed out that this 
condition can in fact be used to reconstruct the $p$-brane action.

It is interesting to speculate on the relationship of the approach in 
this paper to the original formulation of the ``brane scane'' in terms of 
the existence of suitable closed differential forms in the embedding 
superspace from which the Wess-Zumino term in the covariant 
Green-Schwarz $p$-brane action can be constructed \cite{Towns1,AT}. The 
closure of the algebra of worldvolume diffeomorphisms, 
$\kappa$-symmetry transformations and Lorentz transformations 
corresponds to an integrability condition on certain vector fields, 
and this usually has a dual formulation in terms of differential 
forms. The results of this paper are presumably also closely related to the 
geometric interpretation of $\kappa$-symmetry transformations in terms 
of superembeddings \cite{Gates,STV,Sorokin}. 

\vspace{1cm}

\noindent
{\bf Acknowledgement:} I wish to thank Professor P. Townsend for 
pointing out reference \cite{Towns6}.

\section*{Appendix}

In this appendix, the conditions under which the generators of diffeomorphisms 
and $\kappa$-symmetry transformations form a closed algebra under 
commutation are examined.  

Consider first the diffeomorphisms, generated by the right action 
(\ref{diffeo2}). The generator of diffeomorphisms depends on the 
point $Z(\sigma)$ on the embedded worldvolume through the presence of 
$E_i{}^A(Z(\sigma))$ and $\Omega_i{}^{ab}(Z(\sigma))$ (the latter vanishes in 
flat superspace). Thus, if a pair of diffeomorphisms are successively 
applied, the generator of the second diffoemorphism depends on 
the point $Z'$ resulting from the first diffeomorphism:
$$g(Z) \, \rightarrow g(Z) \, e^{i v_1^i(\sigma)(E_i{}^A(Z)T_A + 
\Omega_i{}^{ab}(Z) J_{ab})}\, e^{i 
v_2^j(\sigma) (E_j{}^B(Z')T_B + 
\Omega_j{}^{cd}(Z') J_{cd})}.$$
For infinitesimal diffeomorphisms, the arguments of the exponentials can 
be combined as
\begin{eqnarray*}
 \, & \, & i (v_1 + v_2)^i\, (E_i{}^A T_A + \Omega_i{}^{ab} J_{ab}) \\
 \, & \, & + \frac12 [i v_1^i\, (E_i{}^AT_A + \Omega_i{}^{ab} J_{ab}), 
i  v_2^j\, (E_j{}^BT_B + \Omega_j{}^{cd}J_{cd})]\\
 \, & \, & +  i v_2^j \,( \delta_{v_1}E_j{}^BT_B + 
\delta_{v_1}\Omega_j{}^{cd}J_{cd}) .
\end{eqnarray*}
If the diffeomorphisms are performed in reverse order, the same expression 
with 1 and 2 interchanged results. The commutator of a pair of 
diffeomorphisms is the difference of these expressions. Using the 
result (\ref{X})
and the expression (\ref{Maurer-Cartan}) for the torsion and curvature 
of the coset  
superspace, one finds
$$[\delta_{v_1}, \delta_{v_2}] = \delta_{v_3}$$
with $v_3^i = v_2^j(\partial_j v_1^i) -  v_1^j(\partial_j v_2^i).$

In checking the commutator  of a diffeomorphism and a $\kappa$-symmetry 
transformation, the  variations of the pulled back vielbein and 
connection  are required:
\begin{eqnarray}
\, \delta_{\kappa} E_i{}^a &=& -i E_i{}^{\beta}(\bar{\kappa}P_+C^{-1})^{\alpha} 
i  f_{\alpha \beta}{}^a   -  i 
(\bar{\kappa}P_+C^{-1}  E^M) \Omega_M{}^{cd} E_i{}^b i 
f_{cd \, b}{}^a    \nonumber \\
\, \delta_{\kappa} E_i{}^{\alpha} &=& \partial_i 
(\bar{\kappa}P_+ C^{-1})^{\alpha}
 + i (\bar{\kappa}P_+C^{-1})^{\beta} E_i{}^a i f_{a \beta}{}^{\alpha} 
\nonumber \\
& \, &  + i  (\bar{\kappa} P_+ C^{-1})^{\beta} \Omega_i{}^{ab} i 
f_{ab \, \beta }{}^{\alpha} - i (\bar{\kappa} P_+ C^{-1} 
E^M) \Omega_M{}^{ab} E_i{}^{\beta} i f _{ab \, \beta}{}^{\alpha} 
\nonumber \\
\,\delta_{\kappa} \Omega_i{}^{ab} &=& \partial_i \left(  (\bar{\kappa} P_+ 
C^{-1}E^M) \Omega_M{}^{ab} \right) 
  - i E_i{}^{\beta} 
(\bar{\kappa }P_+ C^{-1})^{\alpha} i f_{\alpha \beta}{}^{ab} \nonumber \\ & \, &
- i
(\bar{\kappa} P_+ C^{-1}E^M) \Omega_M{}^{cd} 
\Omega_i{}^{ef} i f_{cd \, ef}{}^{ab} , 
\label{deltaE}
\end{eqnarray}
where $( \bar{\kappa} P_+ C^{-1}  E^M) = 
(\bar{\kappa} P_+ C^{-1})^{\alpha} E_{\alpha}{}^M,$ and 
where all transformations are evaluated at a point $Z(\sigma)$ on the 
the embedded worldvolume. The results (\ref{deltaE}) are obtained by 
computing
$ g(Z')^{-1} \frac{\partial }{\partial \sigma^i } g(Z') $ 
with $g(Z')$ given by (\ref{kappa}). 
Also required is the $\kappa$-symmetry 
transformation of the projection operator $P_+,$ which is
\begin{equation}
\delta_{\kappa} P_+ =   \frac12 \delta_{\kappa} \Gamma
 = \frac 12 (\delta_{\kappa} E_i{}^{\bar{a}}) 
\Gamma_{\bar{a}} \Gamma^i \Gamma .
\label{deltaP}
\end{equation}
Here, $ \Gamma_{\bar{a}}$ is shorthand for $(1-K)_a{}^b \Gamma_b,$ 
and a sum over $\bar{a}$ is thus a restriction to a sum over the 
spacetime vector degrees of freedom  which are transverse to the worldvolume. 
In computing $\delta_{\kappa}P_+,$
 it is necessary to use the fact that $\Gamma^b 
K_b{}^a = \Gamma^i E_i{}^a$ anticommutes with $\Gamma^c (1 - K)_c{}^d,$ 
and that $\sqrt{\det G}\,
 \epsilon^{i_0 i_1 \cdots i_p}$ transforms as a scalar.

As with the commutator of a pair of diffeomorphisms, 
in computing the commutator of a diffeomorphism and a 
$\kappa$-symmetry transformation, the second transformation acts at a 
transformed point, and one finds
\begin{eqnarray*} 
[\delta_v,\delta_{\kappa}] &=& - [v^i(E_i{}^A T_A + 
\Omega_i{}^{ab}J_{ab}), (\bar{\kappa} P_+ Q) + (\bar{\kappa}P_+ 
C^{-1}E^M) \Omega_M{}^{ab} J_{ab}] \\
& \, &\, - i v^i (\delta_{\kappa} E_i{}^A) T_A - i v^i (\delta_{\kappa} 
\Omega_i{}^{ab})J_{ab} \\
& \, & \, + i \bar{\kappa} (\delta_v P_+)Q + i \bar{\kappa} 
\delta_v (P_+C^{-1}E^M \Omega_M{}^{ab}) 
J_{ab}. 
\end{eqnarray*}
By computing $g(Z') \partial_i g(Z')$ with $g(Z')$ given by (\ref{kappa}), 
the first three terms in this expression are easily shown to be 
$$ - i \partial_i(\bar{\kappa} P_+)Q - i \partial_i((\bar{\kappa} P_+ 
C^{-1}E^M) \Omega_M{}^{ab}) J_{ab}.$$ 
Combining this with the fact that $\delta_v \phi = v^i \partial_i \phi$ for 
worldvolume scalars yields 
$$[\delta_v,\delta_{\kappa}] = \delta_{\kappa'}$$
with $\kappa' = - v^i \partial_i \kappa.$ 

The only difficult step in computing the algebra of right transformations 
 involves a pair of 
$\kappa$-symmetry transformations. This involves
\begin{eqnarray*}
[\delta_{\kappa_1}, \delta_{\kappa_2}] & = & 
[i\bar{\kappa_1}P_+ Q + i (\bar{\kappa}_1 P_+C^{-1}E^M)
 \Omega_M{}^{ab} J_{ab}, i\bar{\kappa_2}P_+ Q \nonumber \\
 & \, & \,+ i 
(\bar{\kappa}_2 P_+C^{-1}E^N) \Omega_N{}^{cd} J_{cd}] 
+ i \bar{\kappa_2} 
(\delta_{\kappa_1} P_+) Q
\nonumber \\
 & \, & \,
 + i \bar{\kappa_2} \delta_{\kappa_1}(  
P_+ C^{-1}E^M \Omega_M{}^{ab}) J_{ab}
 - i \bar{\kappa_1}  
(\delta_{\kappa_2} P_+) Q 
\nonumber \\
& \, & \, - i \bar{\kappa_1} \delta_{\kappa_2}  
(P_+ C^{-1}E^M \Omega_M{}^{ab}) J_{ab}.
\end{eqnarray*}
Computing the commutator and collecting all the terms proportional 
to $J_{ab}$ in a Lorentz transformation $\delta_{L},$
\begin{eqnarray*}
[\delta_{\kappa_1}, \delta_{\kappa_2}] & = & 
2(\bar{\kappa}_2 P_+ \Gamma^a P_{\pm} \kappa_1) P_a  
 \\ & \, & \, 
+ (\bar{\kappa}_2P_+C^{-1}E^M)\Omega_M{}^{ab} 
(\bar{\kappa_1}P_+C^{-1})^{\alpha} i f_{ab \,\alpha }{}^{\beta} 
Q_{\beta}\\ & \, & \, 
 - (\bar{\kappa}_1 P_+C^{-1}E^M)\Omega_M{}^{ab} 
(\bar{\kappa_2}P_+C^{-1})^{\alpha} i f_{ab \,\alpha }{}^{\beta} 
Q_{\beta}  \\ & \, & \, 
+\frac{i}{2} \bar{\kappa}_2 (\delta_{\kappa_1} \Gamma) Q  - 
\frac{i}{2} \bar{\kappa}_1 (\delta_{\kappa_2} \Gamma) Q
+\delta_L,
\end{eqnarray*}
where the $\pm$ in the first term applies to $p$ even/odd, and we have 
used $(C P_+)^T = - CP_{\pm}$ for $P$ even/odd. In fact,
$$ \Gamma^b (1-K)_b{}^a P_{\pm} = P_- \Gamma^b (1-K)_b{}^a,$$
so the first term becomes
$$ 2(\bar{\kappa}_2 P_+ \Gamma^i P_{\pm} \kappa_1) E_i{}^aP_a,$$
which is part of the  generator of a diffeomorphism with parameter
$v^i = -2 i (\bar{\kappa}_2 P_+ \Gamma^i P_{\pm} \kappa_1).$ 
Completing the generator of the diffeomorphism,
\begin{eqnarray*}
[\delta_{\kappa_1}, \delta_{\kappa_2}] & = & \delta_v - 2 
(\bar{\kappa}_2 P_+ \Gamma^i P_{\pm} \kappa_1) 
E_i{}^{\alpha}Q_{\alpha} \\
&\, & + (\bar{\kappa}_2P_+C^{-1}E^M)\Omega_M{}^{ab} 
(\bar{\kappa_1}P_+C^{-1})^{\alpha} i f_{ab \,\alpha }{}^{\beta} 
Q_{\beta}\\
& \, &  - (\bar{\kappa}_1 P_+C^{-1}E^M)\Omega_M{}^{ab} 
(\bar{\kappa_2}P_+C^{-1})^{\alpha} i f_{ab \,\alpha }{}^{\beta} 
Q_{\beta}  \\ & \, & \, 
+\frac{i}{2} \bar{\kappa}_2 (\delta_{\kappa_1} \Gamma) Q  - 
\frac{i}{2} \bar{\kappa}_1 (\delta_{\kappa_2} \Gamma) Q
+\delta_L.
\end{eqnarray*}
The third and fourth terms on the right-hand side  cancel against 
the   $\Omega_M{}^{ab}$ dependent pieces of the last two terms, which are 
obtained from (\ref{deltaE}) and (\ref{deltaP}). 
To show that the remaining $Q$ dependent terms constitute a 
$\kappa$-symmetry transformation, it is only necessary to show that 
the pieces proportional to $P_-Q$ vanish, as a $\kappa$-symmetry transformation 
is generated by $i\bar{\kappa}P_+Q$ (plus a piece which can be 
absorbed into a Lorentz transformation) and so only involves $P_+Q.$
Thus it suffices to prove  
\begin{eqnarray*} 0 &=& - 2 
(\bar{\kappa}_2 P_+ \Gamma^i P_{\pm} \kappa_1) (\bar{E}_iP_- Q) +
(\bar{\kappa}_1P_+ \Gamma_{\bar{a}} E_i) (\bar{\kappa}_2 \Gamma^{\bar{a}} 
\Gamma^i \Gamma P_- Q) \nonumber \\
& \, & - (\bar{\kappa}_2 P_+ \Gamma_{\bar{a}} E_i) (\bar{\kappa}_1 \Gamma^{\bar{a}} 
\Gamma^i \Gamma P_- Q).
\end{eqnarray*}
The  sums over the 
restricted range $\bar{a}$ in the last terms (i.e. over spacetime vector 
indices transverse to the worldvolume, see after equation 
(\ref{deltaP})) can be replaced by 
unrestricted sums by inclusion of appropriate projection operators, and 
the required identity is
\begin{eqnarray}
0 & = &  - 2 
(\overline{\kappa^{\pm}_2}  \Gamma^i \kappa_1^{\pm}) 
(\overline{E^{\mp}_i}Q^-) +
(\overline{\kappa^{\pm}_1} \Gamma_a E_i^{\mp})
 (\overline{\kappa^{\pm}_2} \Gamma^a 
\Gamma^i \Gamma Q^-) \nonumber \\
& \, & - (\overline{\kappa^{\pm}_2} \Gamma_a E_i^{\mp})
 (\overline{\kappa^{\pm}_1} \Gamma^a 
\Gamma^i \Gamma Q^-),
\label{kappacomm}
\end{eqnarray}
where $\theta^{\pm} = P_{\pm} \theta$ and the upper sign in a given term 
in (\ref{kappacomm}) refers to the case $p$ even, while the lower sign 
refers to the case $p$ odd. The distinction between $p$ even and $p$ odd 
arises because for $p$ even
$$ P_{\pm} \Gamma^i P_{\mp} = 0, \,\,\,\,\,\, P_{\pm} \Gamma^{\bar{a}} 
P_{\pm} = 0,$$
while for $p$ odd,
$$ P_{\pm} \Gamma^i P_{\pm} = 0, \,\,\,\,\,\, P_{\pm} \Gamma^{\bar{a}} 
P_{\mp} = 0.$$

For the case $p=1,$  (\ref{kappacomm}) is easily shown to be true 
if the spacetime gamma matrix  identity 
$$ (C \Gamma^a)_{\alpha (\beta}\, (C \Gamma_a)_{\gamma \delta)}$$ applies 
(where the round brackets denote the totally symmetric piece). The gamma 
matrix identity is applied to the second term, and use must be made of the 
fact that 
$$(\overline{\kappa_2^-} \Gamma_j \kappa_1^-)
(\overline{E_i^+} \Gamma^j \Gamma^i \Gamma Q^-) =  
(\overline{\kappa_2^-} \Gamma_j \kappa_1^-)G^{ij}(\overline{E_i^+} \Gamma Q^-) +
(\overline{\kappa_2^-} \Gamma^i \Gamma \kappa_1^-)(\overline{E_i^+}  Q^-)$$
via the $p=1$ version of the identities
\begin{eqnarray}
\, \Gamma^i \, \Gamma & = &  
\frac{i^{[\frac{p+1}{2}]}}{p! } \, \epsilon^{i \,i_1 \cdots i_p}\,
\Gamma_{i_1} \cdots \Gamma_{i_p} \nonumber \\
\, [\Gamma^i, \Gamma^j] \, \Gamma & = & - 2  
\frac{i^{[\frac{p+1}{2}]} }{(p-1)!} 
\, \epsilon^{i\, j\, i_2 \cdots i_p}\, \Gamma_{i_2} \cdots \Gamma_{i_p}.
\label{gammaident}
\end{eqnarray}

For the case $p > 1,$ the situation is a little more complicated. The 
identity (\ref{kappacomm}) can be proved in spacetime dimensions for 
which the gamma matrix identity
$$ 0 = (C \Gamma_{a_1})_{(\alpha \beta} \, (C \Gamma^{a_1 a_2 \cdots 
a_p})_{\gamma \delta)}$$ is true,
where $\Gamma^{a_1 a_2 \cdots a_p}$ is the totally antisymmetric 
product of $p$ gamma matrices. 
To apply the gamma matrix identity, we write
\begin{eqnarray*}
 (\overline{\kappa^{\pm}_1} \Gamma_{\bar{a}} E_i^{\mp})
  (\overline{\kappa^{\pm}_2} \Gamma^{\bar{a}} 
\Gamma^i \Gamma Q^-) & =& i^{[\frac{p+1}{2}]} \frac{1}{p!} 
\, \epsilon^{i \, i_1 \cdots i_p} \\ & \, & \,
(\overline{\kappa^{\pm}_1} \Gamma^a E_i^{\mp})
(\overline{\kappa^{\pm}_2} \Gamma_{a\,  i_1 \cdots i_{p-1}} \, 
\Gamma_{i_p}Q^-).
\end{eqnarray*}
If, {\em in addition}, the condition
\begin{equation}
0 = (P_- \Gamma^i E_i)^{\alpha} = (\Gamma^i E_i^{\mp})^{\alpha}
\label{eqnmotion}
\end{equation}
is satisfied, then the gamma matrix identity 
 (after use of (\ref{gammaident})) yields
\begin{eqnarray*}
0 &=& (\overline{\kappa^{\pm}_1} \Gamma_{\bar{a}} E_i^{\mp}) 
(\overline{\kappa^{\pm}_2} \Gamma^{\bar{a}} 
\Gamma^i \Gamma Q^-) \\
&-& \frac{(p+1)}{p} (\overline{\kappa^{\pm}_2} \Gamma^i \kappa_1^{\pm}) 
(\overline{E_i^{\mp}}Q^-) \\
&-& (\overline{\kappa^{\pm}_2} \Gamma_{\bar{a}} 
E_i^{\mp})((\overline{\kappa^{\pm}_1} \Gamma^{\bar{a}} \Gamma^i \Gamma Q^-) \\
& -&  \frac{(p+1)}{p} (\overline{\kappa^{\pm}_2} \Gamma^i \kappa_1^{\pm}) 
(\overline{E_i^{\mp}}Q^-) \\
& - & \frac{1}{p}  (\overline{\kappa^{\pm}_2} \Gamma_{\bar{a}} E_i^{\mp}) 
(\overline{\kappa^{\pm}_1} \Gamma^{\bar{a}} \Gamma^i \Gamma Q^-) \\
& + & \frac{1}{p}  (\overline{\kappa^{\pm}_1} \Gamma_{\bar{a}} E_i^{\mp}) 
(\overline{\kappa^{\pm}_2} \Gamma^{\bar{a}} \Gamma^i \Gamma Q^-).
\end{eqnarray*}
This is the required result (\ref{kappacomm}). 
Note that the identity (\ref{kappacomm}) is not true if the condition 
is not enforced
(\ref{eqnmotion}), as otherwise there remain terms containing 
$(\Gamma^i E_i^{\pm})^{\alpha}$ which {\em do not} cancel against 
each other.


\begin{thebibliography}{99}
\bibitem{GS}
M. Green and J. Schwarz, Phys. Lett. {\bf 136B} (1984) 367; Nucl. 
Phys. {\bf B243} (1984) 285.

\bibitem{Azc} J.A. Azc\'arraga and J. Lukierski, Phys. Lett. {\bf B113} 
(1982) 170; Phys. Rev. {\bf D28} (1983) 1337.

\bibitem{Siegel} 
W. Siegel, Phys. Lett. {\bf 128B} (1983) 397.

\bibitem{Polch1}
J. Hughes and J. Polchinski, Nucl. Phys. {\bf B278} (1986) 147;

\bibitem{Polch2}
J. Hughes, J. Lui and J. Polchinski,  Phys. Lett. {\bf B180} (1986) 
370.

\bibitem{Mezincescu}
M. Henneaux and L. Mezincescu, Phys. Lett. {\bf B152} (1985) 340.

\bibitem{AT} J.A. de Azc\'arranga and P.K. Townsend, Phys. Rev. {\bf 
62} (1989) 2579.


\bibitem{Tseytlin2}
R.R. Metsaev and A.A. Tseytlin,  Phys. Lett. {\bf B436}, (1998) 281.

\bibitem{STV}
D. Sorokin, V. Tkach and D.V. Volkov, Mod. Phys. Lett. {\bf A4} 
(1989) 901.

\bibitem{Gates} S.J. Gates Jr. and H. Nishino, Class. Quantum Grav. 
{\bf 3} (1986) 391; R. Brooks, F. Muhammed and S.J. Gates Jr,  
Class. Quantum Grav. {\bf 3} (1986) 745.

\bibitem{Sorokin}
D. Sorokin, {\em Superbranes and Superembeddings,} hep-th/9906142.

\bibitem{GSW}
M. Green, J. Schwarz and E. Witten, {\em Superstring Theory, Volume 
1}, Cambridge University Press (1987), Chapter 5.

\bibitem{Salam}
A. Salam and J. Strathdee, Nucl. Phys. {\bf B76} (1974) 477.

\bibitem{Towns6} J.A. de Azc\'arraga, J.M. Izquierdo and P.K. 
Townsend, Phys. Lett. {\bf B267} (1991) 366.

\bibitem{Bars}I. Bars, C. Deliduman and D. Minic, Phys. Rev {\bf D59} 
(1999) 125004; Phys. Lett. {\bf B457} (1999) 275.

\bibitem{Witten1}
E. Witten, Nucl. Phys. {\bf B266} (1986) 245.

\bibitem{Gris1}
M.T. Grisaru, P. Howe, L. Mezincescu, B.E.W. Nilsson and P.K. 
Townsend, Phys. Lett. {\bf 162B} (1985) 116.

\bibitem{Towns1} A. Achucarro, J. Evans, P. Townsend and D. Wiltshire, 
Phys. Lett. {\bf 198B} (1987) 441.



\bibitem{Strathdee}
A. Salam and J. Strathdee, Ann. Phys. {\bf 141} (1982) 316.

\bibitem{Tseytlin1}
R.R. Metsaev and A.A. Tseytlin,  Nucl. Phys. {\bf B533} (1998) 109.


\bibitem{Plefka}
B. de Wit, K. Peeters and J.C. Plefka,  Nucl. Phys. {\bf B532} (1998) 
99.

\bibitem{Strathdee2}
A. Salam and J. Strathdee, Nucl. Phys. {\bf B76} (1974) 477.

\bibitem{Towns2}
E. Bergshoeff, E. Sezgin and P.K. Townsend, Phys. Lett. {\bf B189} 
(1987) 75; Ann. Phys.  {\bf 185} (1988) 330.


\bibitem{Plefka2}
B. de Wit, K. Peeters, J.C. Plefka and A. Sevrin,  Phys .Lett. {\bf 
B443} (1998) 
153.

\bibitem{Claus}
P. Claus, Phys. Rev. {\bf D59} (1999) 066003.

\bibitem{Nicolai}
See, for example, section 2.2 of the lectures by H. Nicolai in {\em 
 Recent Aspects of Quantum 
Fields, }  Proceedings, Schladming 1991, eds. H. Mitter and H. 
Gausterer, Springer Verlag, 1991.

\bibitem{WZconsistency} J. Wess and B. Zumino, Phys. Lett. {\bf 37B} 
(1971) 95. 

\bibitem{Schwarzetal}
M. Aganagic, C. Popescu and J.H. Schwarz, Phys. Lett. {\bf B393} 
(1997) 311; Nucl. Phys. {\bf B495} (1997) 99.

\bibitem{wess2}
J. Wess and B. Zumino, Phys. Lett. {\bf 79B} (1978) 394.




\end{thebibliography}
\end{document}